\documentclass[conference,10.pt]{IEEEtran}
\usepackage{verbatim}  
\usepackage{epsf}
\usepackage{epsfig}
\usepackage{epstopdf}
\usepackage{enumerate}
\usepackage{times}
\usepackage{amsmath}
\usepackage{amssymb}
\usepackage{cite}

\IEEEoverridecommandlockouts

\usepackage{algorithm}
\usepackage{algpseudocode}

\usepackage[hidelinks,bookmarks=false]{hyperref}

\title{Updating Beamformers to Respond to Changes in Users}

\author{\IEEEauthorblockN{Mostafa Medra$^\ast$ \quad Andrew W. Eckford$^\dagger$ \quad Raviraj Adve$^\ast$}
\IEEEauthorblockA{$^\ast$ Department of Electrical and Computer Engineering, University of Toronto, ON, Canada. \\
$^\dagger$ Department of Electrical Engineering and Computer Science, York University, Toronto, ON, Canada.}
\thanks{This research was supported by TELUS Canada and the Natural Sciences
and Engineering Research Council (NSERC) of Canada.}
}

\begin{document}
\maketitle

\begin{abstract}

We consider a multi-user multiple-input single-output downlink system that provides each user with a prespecified level of quality-of-service. The base station (BS) designs the beamformers so that each user receives a certain signal-to-interference-and-noise ratio (SINR). In contrast to most of the available literature in the beamforming field, we focus on the required modifications when the system changes. We specifically study three cases: (i) user entering the system, (ii) user leaving the system, and (iii) a change in the SINR target. We do so in order to avoid designing the entire system from scratch for every change in the requirements.  In each of the three cases, we describe the modifications required to the beamforming directions and the power loading. We consider maximum ratio transmission (MRT), zero-forcing (ZF) and the optimal beamformers. The proposed modifications provide performance that is either exact or very close to that obtained when we redesign the entire system, while having much lower computational cost.

\end{abstract}

\section{Introduction}

When the base station (BS) is equipped with multiple antennas, beamforming methods can be used to serve multiple users simultaneously \cite{Writingondirtypaper,Avectorperturbationtechnique2,Precodinginmultiantenna,SymbollevelandMulticast,Rate_Splitting_Bridging,ShiftingtheMIMO}. Among these beamforming algorithms, dirty paper coding \cite{Writingondirtypaper} is capacity achieving~\cite{TheCapacityRegion}. However, it is of high computational cost and linear beamforming techniques are used instead; e.g., \cite{Reference2,Jointoptimal,Linearprecoding,Iterativemultiuseruplink,OptimalMultiuserTransmit}. The simplest beamforming method is the maximum ratio transmission (MRT) \cite{Maximumratiotransmission}, in which the beamforming directions are chosen to match the individual users' channels. However, such a technique does not take into account the interference between the users and can only provide good performance with a large number of antennas. Another appealing method is the zero-forcing (ZF) technique \cite{Zeroforcingmethods}, in which the beamformers are chosen to maximize the received signal, and be in the null space of the channels of other users. Both the MRT and ZF directions are then provided with specific power loads so that each user receives the required signal-to-interference-and-noise ratio (SINR).  In general the problem of minimizing the transmission power under SINR constraints can be formulated as a convex problem and the optimum solution can be written in closed-form expressions~\cite{Jointoptimal,Reference2,Solutionofthemultiuser,OptimalMultiuserTransmit}.

Most of the literature in beamforming techniques has focused on designing the entire system from scratch. Such an approach waste computational resources requiring a complete redesign for every change. In realistic situations, it is quite possible that a new user enters the system while other users' channels are still considered good estimates of the true channels. A user may also leave the system, or require a different rate to support a different application.

The question we answer here is how to modify the already computed beamformers to accommodate for the system change without the need to redesign the entire system, and to do so using few computational resources. One example for this kind of update is presented in \cite{FastMatrixInversion} where the authors used the mathematical relationship that relates a partitioned matrix to its blocks \cite{duncan1944lxxviii} to update the ZF beamforming without having to do a matrix inversion again. Another example is presented in \cite{RobustUpdateAlgorithms}, where the authors updated an extension of the QR decomposition to obtain the new beamformers.

In this paper, we introduce the system model and provide the closed form expressions for the MRT, ZF, and the optimal beamformers in Section~\ref{Sec_system_model}. In Section~\ref{Sec_math}, we review and introduce some important mathematical relationships that will be used to describe the required beamformers' changes. In Section~\ref{dir_changes}, we explain the required modifications in the beamforming directions and in Section~\ref{PL_changes} we explain the modifications for the power loading for each of the beamforming schemes in each scenario. We provide exact expressions for the modifications in the case of MRT and ZF directions. In the optimal beamformers case, we provide approximations that perform very close compared to the optimal performance.

\section{System model and design approach}\label{Sec_system_model}

We consider a narrowband  multiple-input single-output (MISO) downlink system with an $N_t$-antenna BS and $K$ single-antenna users. The transmitted signal  $\mathbf{x}$ is constructed using linear beamforming such that $\mathbf{x}= \sum_{k=1}^K\mathbf{w}_k s_k$, where $s_k$ is the normalized data symbol for user $k$, and $\mathbf{w}_k$ is the associated beamformer. The received signal  at user $k$ is
\begin{equation}\label{rcvd_sig}
    y_k= \mathbf{h}_k^H \mathbf{w}_k s_k + \textstyle\sum_{j \neq k}\mathbf{h}_k^H \mathbf{w}_j s_j + n_k,
\end{equation}
where $\mathbf{h}_k^H$ denotes the channel between the BS and receiver $k$, and $n_k$ represents the additive zero-mean circular complex Gaussian noise at that user.

In the problems we will consider, the operating rate for each user can be translated to a signal-to-interference-and-noise ratio (SINR) target, $\gamma_k$, which takes the form
\begin{equation}\
    \text{SINR}_k =  \frac{\mathbf{h}_k^H \mathbf{w}_k \mathbf{w}_k^H \mathbf{h}_k}{\mathbf{h}_k^H  (\sum_{j \neq k}\mathbf{w}_j \mathbf{w}_j^H) \mathbf{h}_k + \sigma_k^2} \geq \gamma_k, \\
\end{equation}
or equivalently $\mathbf{h}_{k}^H \mathbf{Q}_{k} \mathbf{h}_{k} - \sigma_k^2 \geq 0$, where
\begin{equation}
\mathbf{Q}_{k}=\mathbf{w}_k \mathbf{w}_k^H/\gamma_k-\sum_{j \neq k}\mathbf{w}_j \mathbf{w}_j^H,
\end{equation}
and $\sigma_k^2$ is the noise variance at receiver $k$.

If we denote the transmitted signal from the $i$th antenna  by $x_i$, then the average transmitted  power from the BS can be written as $\textstyle\sum_{i=1}^{N_t} E |x_i|^2 $. In the case of zero-mean, independent data symbols of normalized power, this  average transmitted  power becomes $\textstyle\sum_{k=1}^K  \mathbf{w}_k^H \mathbf{w}_k$.

In this paper we consider three types of beamforming directions; MRT, ZF, and the optimal beamformers. For a given set of SINR targets, $\{\gamma_k\}$, the minimum power required to achieve those SINR values can be obtained by considering the following problem
\begin{subequations}\label{QoS}
\begin{align}
      \min_{\substack{\mathbf{w}_k}} \quad  & \textstyle\sum_{k} \mathbf{w}_k^H \mathbf{w}_k \\
      &  \mathbf{h}_k^H \mathbf{Q}_k \mathbf{h}_k - \sigma_k^2 \geq 0. \label{sinr1}
      \end{align}
\end{subequations}
Since the problem in \eqref{QoS} can be directly transformed into a convex problem, the resulting beamformers are optimal in terms of the transmission power  \cite{Reference2}.  To obtain such beamformers, if we let $\nu_k$ denote the dual variable of the constraint in \eqref{sinr1}, then the  dual variables $\{\nu_k\}$ should satisfy the fixed-point equations \cite{Reference2}
\begin{equation}\label{nu}
\nu_k^{-1}  = \mathbf{h}_{k}^H \Bigl(\mathbf{I}+\textstyle\sum_{j} \nu_j \mathbf{h}_{j} \mathbf{h}_{j}^H \Bigr)^{-1} \mathbf{h}_{k}  \Bigl(1+\frac{1}{\gamma_k} \Bigr).
\end{equation}
From the KKT conditions, the beamforming directions can be written as
\begin{equation}\label{w_eqn}
\mathbf{w}_k =\Biggl( \frac{\nu_k}{\gamma_k}\mathbf{h}_{k} \mathbf{h}_{k}^H-\sum_{j\neq k} \nu_j \mathbf{h}_{j} \mathbf{h}_{j}^H \Biggr)\mathbf{w}_k,
\end{equation}

Accordingly, by solving the fixed-point equations in \eqref{nu}, we obtain the dual variables $\{\nu_k\}$ by which we can solve the eigen equations in \eqref{w_eqn} to obtain the beamforming directions. The $K$ unknown amplitude squares $\beta_k= \|\mathbf{w}_k\|^2$ are calculated from the $K$ linear equations that are derived from the fact that at optimality all the constraints in \eqref{sinr1} hold with equality~\cite{Reference2}.
If we  define $\boldsymbol{\beta}=[\beta_1, \beta_2,..., \beta_K]^T$, $\boldsymbol{\sigma}^2=[\sigma_{1}^2, \sigma_{2}^2,..., \sigma_{K}^2]^T$, and the matrix $\mathbf{A}$ such that $\mathbf{[A]}_{ii}= | \mathbf{h}_{i}^H {\mathbf{u}}_i |^2/\gamma_i$, and $\mathbf{[A]}_{ij}= - | \mathbf{h}_{i}^H {\mathbf{u}}_j |^2$,  $\forall i \neq j$, then  the set of linear equations can be written as
\begin{equation}\label{A_eqn}
   \mathbf{A} \boldsymbol{\beta} =\boldsymbol{\sigma}^2,
\end{equation}
resulting in $\boldsymbol{\beta} = \mathbf{A}^{-1} \boldsymbol{\sigma}^2.$

The beamformer of user $k$ can be obtained by normalizing $\mathbf{h}_{k}$ in the MRT case, from the $k$th column of $\mathbf{H} (\mathbf{H}^H \mathbf{H})^{-1}$ in the ZF case, and optimally by solving \eqref{nu} and \eqref{w_eqn} in the perfect CSI case. The power loading for the three cases is done by solving \eqref{A_eqn}.

\section{Mathematical formulae}\label{Sec_math}

In this section we will review and derive some important formulae that will be used in this paper.

\subsection{Block matrix inversion \cite{duncan1944lxxviii}}\label{ZF1}

We can partition a matrix into four blocks so that its inverse is related to the blocks as follows
\begin{equation}\label{block_mat_inv}
\left(
  \begin{array}{cc}
    \mathbf{A} & \mathbf{B} \\
    \mathbf{C} & \mathbf{D} \\
  \end{array}
\right)^{-1}= \left(
                \begin{array}{cc}
                  \mathbf{A}^{-1}+\mathbf{A}^{-1} \mathbf{B} \mathbf{E} \mathbf{C} \mathbf{A}^{-1} & -\mathbf{A}^{-1} \mathbf{B} \mathbf{E} \\
                  -\mathbf{E} \mathbf{C} \mathbf{A}^{-1} & \mathbf{E} \\
                \end{array}
              \right),
\end{equation}
where $\mathbf{E}=( \mathbf{D}-\mathbf{C} \mathbf{A}^{-1} \mathbf{B})^{-1}.$
When we have the $\mathbf{A}^{-1}$ and $\mathbf{E}$ blocks  already computed, we can find the inverse for the augmented matrix  using matrix multiplications as shown above without the need to actually do the inversion.

The other case that is useful in this paper is when we have the matrix inversion of the augmented matrix and we want to obtain  $\mathbf{A}^{-1}$ from the already computed blocks. In this case we realize that $\mathbf{A}^{-1}=(\mathbf{A}^{-1}+\mathbf{A}^{-1} \mathbf{B} \mathbf{E} \mathbf{C}  \mathbf{A}^{-1}) - (\mathbf{A}^{-1} \mathbf{B} \mathbf{E}) \mathbf{E}^{-1} (\mathbf{E} \mathbf{C} \mathbf{A}^{-1}),$ where all the required blocks (written in parenthesis) are already computed. While these relationships can be very useful, they only work on square matrices for $\mathbf{A}$ and $\mathbf{D}$.

\subsection{Alternate way to update the matrix inversion}\label{ZF2}

In this section, we will derive an alternate way to update the matrix inverse directly.
Assume that we calculated  $\mathbf{G}=\mathbf{H} (\mathbf{H}^H \mathbf{H})^{-1}$; i.e., $\mathbf{G}^H \mathbf{H}=\mathbf{I}$. Now we will show how to calculate $\mathbf{G}_{K-1}={\mathbf{H}}_{K-1}^{\dag}$ given that we already calculated  $\mathbf{G}$, where $\mathbf{H}_{K-1}$ is equal to $\mathbf{H}$, but with the last column removed. We know that each column of $\mathbf{G}$ is orthogonal to all the columns of $\mathbf{H}$  except the corresponding column. Let us call the $j$th column of $\mathbf{G}$ by $\mathbf{g}_j$ and the $k$th column of $\mathbf{H}$ by $\mathbf{h}_k$. The new $j$th column in $\mathbf{G}_{K-1}$ should also be orthogonal to all the columns in $\mathbf{H}_{K-1}$ except the $j$th column and is no more required to be orthogonal to the $K$th eliminated column. The component in the direction of $\mathbf{h}_k$, but not in the space of other $\mathbf{h}_j, j \neq K$ is $\mathbf{g}_K$. Accordingly, the general form of the new $j$th column is  $\hat{\mathbf{g}}_j=a \mathbf{g}_j + b \mathbf{g}_K$, where $a$ and $b$ are constants. We can check that this general form is orthogonal to all the remaining columns in $\mathbf{H}_{K-1}$ except for the corresponding column $j$; i.e., $\mathbf{h}_i^H (a \mathbf{g}_j + b \mathbf{g}_K)=a \mathbf{h}_i^H \mathbf{g}_j + b \mathbf{h}_i^H \mathbf{g}_K=0, \forall i \neq j, K$. Now to find the constants $a, b$ in $\hat{\mathbf{g}}_j$, we can formulate that problem as
\begin{subequations}\label{ch_inv}
\begin{align}
    \min_{\substack{a,b}} \    \quad &  \| a \mathbf{g}_j + b \mathbf{g}_K  \|  \\
   \text{s.t.} \quad   & (a \mathbf{g}_j + b \mathbf{g}_K)^H \mathbf{h}_j=1.
       \end{align}
\end{subequations}
Note that obtaining the vector of the minimum norm that satisfies $\hat{\mathbf{g}}_j^H \mathbf{h}_j=1$ is the same as maximizing the inner product between $\hat{\mathbf{g}}_j$ and $\mathbf{h}_j$ under a constraint on the norm of  $\hat{\mathbf{g}}_j$. In both cases, the resulting constants $a$ and $b$ minimize the angle between both vectors. Accordingly, $\hat{\mathbf{g}}_j$ provides the maximum inner product and zero-interference to the other users, which is what a ZF direction does.
Since $\mathbf{g}_K^H$, and $\mathbf{h}_j$ are orthogonal, the equality constraint is reduced to $a=1$ and we are left with the unconstraint minimization problem $\|  \mathbf{g}_j + b \mathbf{g}_K  \| $ which can be easily solved to get $b=- \frac{\mathbf{g}_K^H \mathbf{g}_j }{\mathbf{g}_K^H \mathbf{g}_K}$. This operation is done for each vector $\hat{\mathbf{g}}_j, \forall j \neq K$.

Now for the other possible situation. Assume that we calculated  $\mathbf{G}=\mathbf{H} (\mathbf{H}^H \mathbf{H})^{-1}$ and want to calculate $\mathbf{G}_{K+1}={\mathbf{H}}_{K+1}^{\dag}$, where ${\mathbf{H}}_{K+1}$ is the matrix $\mathbf{H}_K$ concatenated with an extra column $\mathbf{h}_{K+1}$.  If we let $\hat{\mathbf{g}}_j$ denote the $j$th column of the matrix $\mathbf{G}_{K+1}$, then we can obtain $\hat{\mathbf{g}}_{K+1}$ as
$\hat{\mathbf{g}}_{K+1} = c(\mathbf{h}_{K+1} - \mathbf{H} (\mathbf{H}^H \mathbf{H})^{-1} \mathbf{H}^H \mathbf{h}_{K+1}$), where $c$ is a scalar chosen such that $\hat{\mathbf{g}}_{K+1}^H \mathbf{h}_{K+1}=1$. This is true as the matrix  $\mathbf{H} (\mathbf{H}^H \mathbf{H})^{-1} \mathbf{H}^H$ projects $\mathbf{h}_{K+1}$ on its subspace, and the subtraction ensures that $\hat{\mathbf{g}}_{K+1}$ is now orthogonal to the space of $\mathbf{H}$. We can simplify that to
$$\hat{\mathbf{g}}_{K+1} = c(\mathbf{h}_{K+1} -\mathbf{G} \mathbf{H}^H \mathbf{h}_{K+1}).$$
Now to obtain $\hat{\mathbf{g}}_j$, we remove component from $\mathbf{g}_j$ in the direction of $\hat{\mathbf{g}}_{K+1}$ so that $\hat{\mathbf{g}}_j$ is orthogonal to the new vector $\mathbf{h}_{K+1}$. (Note that $\mathbf{g}_j$  is already orthogonal to the other channel vectors except fot the $j$th one.) If we let $\hat{\mathbf{g}}_j = \mathbf{g}_j +b \hat{\mathbf{g}}_{K+1}$, then $\mathbf{h}_{K+1}^H \hat{\mathbf{g}}_j=0$ results in $b= - (\mathbf{h}_{K+1}^H \mathbf{g}_j / \mathbf{h}_{K+1}^H \hat{\mathbf{g}}_{K+1})$. Accordingly,
$$\hat{\mathbf{g}}_j = \mathbf{g}_j - (\mathbf{h}_{K+1}^H \mathbf{g}_j / \mathbf{h}_{K+1}^H \hat{\mathbf{g}}_{K+1})  \hat{\mathbf{g}}_{K+1}.$$
This operation is done for each vector $\hat{\mathbf{g}}_j, \forall j \neq K$.
%

\subsection{Matrix inversion rank-one update \cite{sherman1949adjustment}}\label{rank_one_update}

If we already computed $\mathbf{A}^{-1}$, then  a rank-one update on this inverse, $(\mathbf{A}+ \mathbf{v} \mathbf{v}^H )^{-1}$, can be written as

$$(\mathbf{A}+ \mathbf{v} \mathbf{v}^H )^{-1}=  \mathbf{A}^{-1} - \mathbf{A}^{-1}\mathbf{v} \mathbf{v}^H  \mathbf{A}^{-1}   /(1+  \mathbf{v}^H \mathbf{A}^{-1} \mathbf{v}).$$

\subsection{Required modifications}

In this section, we table the required modifications for each of the possible cases under test. We use here no change (NC) or change (C) to indicate whether a major modification is required or not.
In Table~\ref{table_dir} we summarize the modifications for each beamforming direction, and in  Table~\ref{table_PL} we summarize those for the power loading (PL) for each of the possible cases under test.

\begin{table}
\begin{center}
\caption{Required modifications for the beamforming directions}
\begin{tabular}{|c|c|c|c|}
  \hline
   & MRT directions  & ZF directions &  Optimal directions \\ \hline
  User out & NC & C & C \\ \hline
  User in & NC & C & C \\ \hline
  $ \gamma$ change & NC & NC & C \\
  \hline
\end{tabular}
\label{table_dir}
\end{center}
\end{table}

\begin{table}
\begin{center}
\caption{Required modifications for the power loading}
\begin{tabular}{|c|c|c|c|}
  \hline
   & MRT PL  & ZF PL &  Optimal PL \\ \hline
  User out & C & NC & C \\ \hline
  User in & C & NC & C \\ \hline
  $ \gamma$  change & C & NC & C \\
  \hline
\end{tabular}
\label{table_PL}
\end{center}
\end{table}
%

\section{Changes in beamforming directions}\label{dir_changes}

Since the MRT beamformers for user $k$ are independent of other users and the SINR targets, we will focus on the ZF and the optimal beamformers.

\subsection{ZF beamforming}

The directions of the ZF beamforming can be obtained by normalizing the columns of the matrix $\mathbf{H} (\mathbf{H}^H \mathbf{H})^{-1}$. We can apply the modifications using two approaches. We can update the square matrix inversion $(\mathbf{H}^H \mathbf{H})^{-1}$ by using the formulae in \eqref{ZF1}, then obtain the beamforming directions by multiplying with the new channel matrix. When a user enters the system whose channel is $\mathbf{h}_{K+1}$, so that the channel matrix is $\mathbf{H}_{K+1}$, then $\mathbf{H}_{K+1}^H \mathbf{H}_{K+1}$ can be partitioned as shown in \eqref{ZF1} such that $\mathbf{A}=\mathbf{H}_{K}^H \mathbf{H}_{K}$, the matrices $\mathbf{B}=\mathbf{C}^H = \mathbf{H}_{K}^H \mathbf{h}_{K+1}$ are now vectors, and $\mathbf{E}=\mathbf{h}_{K+1}^H \mathbf{h}_{K+1}$ is a scalar. Accordingly, given the inversion of $\mathbf{A}$, we can find $(\mathbf{H}_{K+1}^H \mathbf{H}_{K+1})^{-1}$ using computations in the order of $\mathcal{O}(K^2)$. After that, the multiplications in $\mathbf{H}_{K+1} (\mathbf{H}_{K+1}^H \mathbf{H}_{K+1})^{-1}$ will cost $\mathcal{O}(N_t K^2)$ operations. This approach was presented in \cite{FastMatrixInversion}. We can derive an analogous steps for the case of a user leaving the system, resulting of a similar computational cost of $\mathcal{O}(N_t K^2)$.

As we have shown, using  the mathematical relationships in \eqref{ZF1} provides an exact update for the ZF beamforming. However, this method requires the knowledge of $(\mathbf{H}^H \mathbf{H})^{-1}$. While the storage of such a matrix can be simple as its dimensions are only $K \times K$, there are recent researches on how to obtain an approximate ZF directions using a truncated polynomial expansion approach to reduce the computational complexity; e.g., \cite{Efficientlinearprecoding}. In that case, the matrix inversion is not available. Instead, the ZF directions can be directly updated using the formulae in \eqref{ZF2} which only require the beamforming directions. The computational cost for both cases of a user entering or leaving is only $\mathcal{O}(N_t K)$. Accordingly, using \eqref{ZF2} provides the same update with lower computational cost, in addition to relaxing the memory requirements.


\subsection{Optimal beamforming}

The exact solution of the optimal beamformers requires solving the problem in \eqref{QoS} whenever any change happens. However, we will show here very efficient approximations of almost negligible degradation in performance to update the beamforming directions.

When a user $(K)$ enters the system, we start by finding the value of $\nu_K$ that satisfies \eqref{nu}. Since $\nu_K$ enters linearly in the matrix inversion, it can be efficiently updated using \eqref{rank_one_update}. That results in
\begin{equation}\label{nu_1}
   \nu_K= \gamma_K/\mathbf{h}_{K}^H \Bigl(\mathbf{I}+\textstyle\sum_{j\neq K} \nu_j \mathbf{h}_{j} \mathbf{h}_{j}^H \Bigr)^{-1}  \mathbf{h}_K.
\end{equation}
Since \eqref{nu} is in the form of a fixed-point equation, to obtain the optimum values we have to iterate on the different $\nu_j$ till convergence. However, we terminate at that step and use the previously calculated $\nu_j$ and the new $\nu_K$ to obtain the beamforming vectors from \eqref{w_eqn}. The reason behind that choice is that when the number of antennas is large and the channels become more orthogonal to each other, the values of $\nu_j$ are less effected by the new user and can provide beamformers that are quite close to the optimal performance.  Since the matrix inversion $(\mathbf{I}+\textstyle\sum_{j\neq K} \nu_j \mathbf{h}_{j} \mathbf{h}_{j}^H)^{-1}$ is already available, the computational cost for obtaining  $\nu_K$ is $\mathcal{O}(N_t^2)$. We also need to update the matrix inversion to obtain $(\mathbf{I}+\textstyle\sum_{j} \nu_j \mathbf{h}_{j} \mathbf{h}_{j}^H)^{-1}$ using \eqref{rank_one_update} which also requires $\mathcal{O}(N_t^2)$. In \cite{OffsetBasedBeamforming}, the authors used the fact that when the channels are almost orthogonal, then the terms $\textstyle\sum_{j\neq K} \nu_j \mathbf{h}_{j} \mathbf{h}_{j}^H$ can be treated as an eigen decomposition, and that reduces \eqref{nu_1} to
\begin{equation}\label{nu_2}
   \nu_K \approx \gamma_K/\mathbf{h}_{K}^H \mathbf{h}_K.
\end{equation}
The approximation in \eqref{nu_2} requires $\mathcal{O}(N_t)$ operations instead of $\mathcal{O}(N_t^2)$. This reduction is significant in massive MISO systems where the value of $N_t$ is typically large.

When a user exits the system, we simply set its corresponding $\nu_K$ to zero, and, using arguments similar to above, we will not update the other $\nu_j$'s. Another way to look at this is to consider the user of a zero-rate; $\gamma_K=0$. Then using \eqref{nu_1} or \eqref{nu_2} we obtain $\nu_K=0$. Accordingly, no computations are involved in this case.

When a user has a new SINR target $\hat{\gamma}_K$, we modify that user's $\nu_K$, denoted $\hat{\nu}_K$, such that the equation in \eqref{nu} holds. Using Section~\ref{rank_one_update}, we can evaluate $\hat{\nu}_K$ using
$$\hat{\nu}_K= \hat{\gamma}_k/\mathbf{h}_{k}^H (\mathbf{I}+\textstyle\sum_{j\neq K} \nu_j \mathbf{h}_{j} \mathbf{h}_{j}^H)^{-1}  \mathbf{h}_{k} - \hat{\gamma}_k \nu_k.$$
Again, we will not update the other $\nu_j$'s. The approximation corresponding to \eqref{nu_2} will result in $\hat{\nu}_K \approx \hat{\gamma}_K/\mathbf{h}_{K}^H \mathbf{h}_K.$

After we update the set of $\{\nu_k\}$, we evaluate the beamforming directions from \eqref{w_eqn}. Since the matrices that require the eigen decomposition are already factorized, the beamforming directions can be efficiently evaluated using power iterations in $\mathcal{O}(N_t K^2)$ operations.

\section{Changes in power loading}\label{PL_changes}

Once we modify the beamforming directions, we have to formulate the matrix $\mathbf{A}$ to obtain the power loading from \eqref{A_eqn}. Since in the ZF case, the matrix $\mathbf{A}$ is diagonal, the power loading for each user is decoupled from the other users and can be obtained directly as $\beta_k=\gamma_k \sigma_k^2/ | \mathbf{h}_{k}^H \mathbf{u}_k|^2$. Accordingly, we will focus on the power loading of  MRT and the optimal case only.

\subsection{MRT power loading}

Since the MRT directions are the same for any change, the entries of the matrix $\mathbf{A}$ of the current users will remain unchanged. When we have an SINR change for user $K$, then only the entry $(K,K)$ will change in matrix $\mathbf{A}$ and the inversion can be updated using the rank-one update procedure in \eqref{rank_one_update}. The cost of that update is $\mathcal{O}(K^2)$. When a user enters/exits the system, the matrix $\mathbf{A}$ would have an extra/less column and row. In this case, the matrix update can be done using the formulae in \eqref{ZF1}, where the blocks $\mathbf{B}$ and $\mathbf{C}$ will correspond to vectors and block $\mathbf{D}$ is a scalar, which simplifies the computations. A similar cost of $\mathcal{O}(K^2)$ is required.

\subsection{Optimal power loading}

In general, the directions of $\mathbf{u}_j$  are modified when any change happens in the system. Accordingly, the matrix $\mathbf{A}$ needs to be recalculated and inverted to obtain the power loading. The matrix inversion requires $\mathcal{O}(K^3)$ operations and calculating the matrix entries requires $\mathcal{O}(N_t K^2)$ operations.

\section{Simulation results}

In this section, we will focus on the optimal beamformers to show how efficient the proposed approximations are. For the ZF and MRT beamformers, the proposed modifications result in exactly the original beamformers and power loading. Accordingly, there is no performance loss.

To illustrate the performance of the proposed algorithms we consider a system consisting of a BS that has $N_t$ antennas and serves $K$ single-antenna users. The users are uniformly distributed in a circle of radius 0.75km around the BS, and the BS height is 25m. We assume a large scale fading model described with a path-loss exponent of 3.52 and log-normal shadow fading with 8dB standard deviation. The small scale fading is modelled using the standard i.i.d. Rayleigh model. We assume that each user has a signal sensitivity of -90dBm, and we will consider this power as the noise power. 

\begin{figure}
\begin{center}
    \epsfysize= 2.8in
     \epsffile{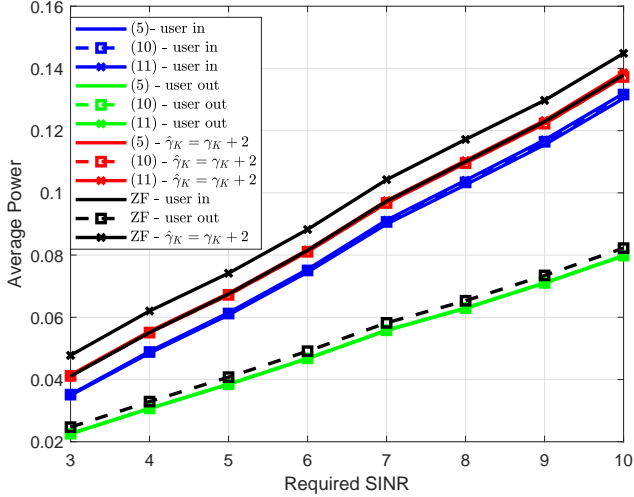}
\caption{Average transmitted power versus the SINR target for $K=4$ users and $N_t=8$ antennas.}\label{fig1}
\end{center}
\end{figure}

\begin{figure}
\begin{center}
    \epsfysize= 2.8in
     \epsffile{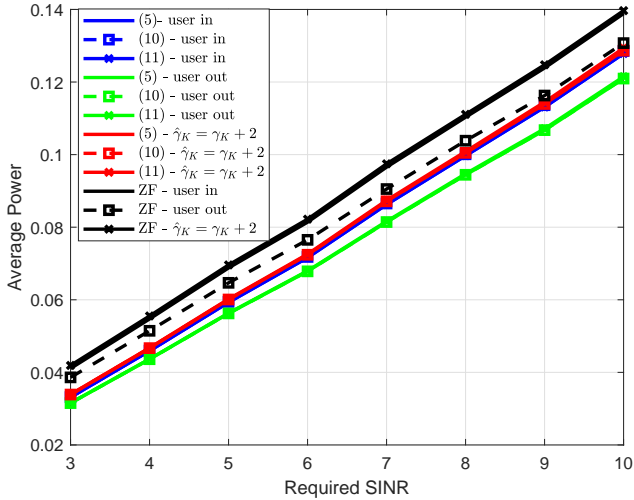}
\caption{Average transmitted power versus the SINR target for $K=32$ users and $N_t=64$ antennas.}\label{fig2}
\end{center}
\end{figure}

We consider the cases of a user entering the system, user leaving the system and an SINR change for a certain user.
For a given $K$, we assume that we have the beamformer design for $K-1$ users, then the $K$th user enters the system for the ``user-in" case. For the ``user-out" case, we assume that we have the beamformer designs for the $K$ users, then user $K$ leaves the system to have a reduced system of $K-1$ users only. For the $\gamma$ change case, we assume that the SINR of user $K$ increases such that $\hat{\gamma}_K=\gamma_K+2$. The main difference between the approximations of the optimal algorithm is how to obtain $\nu_j$. The exact value is obtained using \eqref{nu}, one approximation is by using \eqref{nu_1}, and a simpler approximation using \eqref{nu_2}. To compare between the exact and the approximated algorithms, we plot the resulting average transmitted power versus the required SINR target $\gamma_k$ for two different scenarios. In Fig.~\ref{fig1}, we plot for $N_t=8$ and $K=4$, and in Fig.~\ref{fig2} we plot for $N_t=64$ and $K=32$. As we can see, the suggested approximations are quite effective even for a low number of antennas. We also plotted the ZF beamforming as a reference. As expected, the optimal algorithm performs better than the ZF, the ``user-out" case requires less power than the ``user-in" case for accommodating the extra user, and the ``$\gamma$ change" case requires the most power due to the higher SINR requirement of user $K$. The negligible performance loss in terms of slightly higher power (only in the case of the optimal algorithms) suggests that the proposed beamforming updates can be quite effective in the beamforming field.

\section{Conclusion}
In this paper we addressed the problem of modifying the designed beamformers when the system changes. We investigated three beamforming designs;  maximum ratio transmission (MRT), zero-forcing (ZF), and the optimal beamformers. We examined the cases in which a user enters or leaves the system and the case of an SINR change. We explained the required modifications in terms of beamforming directions and the power loading for each case. For the MRT and ZF case, we showed that we can modify the system to have the exact performance of a complete new design. We derived mathematical relationships that reduced the amount of computations required to update the ZF directions. In the optimal beamforming case, we provided efficient approximations that incurred almost no performance degradation. The reductions in computational complexities are shown to be significant.

\end{document}